\documentclass[fleqn,usenatbib]{mnras}
\pdfoutput=1

\usepackage{newtxtext,newtxmath}

\usepackage[T1]{fontenc}
\usepackage{ae,aecompl}
\usepackage{graphicx}   % Including figure files
\usepackage{amsmath}    % Advanced maths commands
\usepackage{amssymb}    % Extra maths symbols

\usepackage{makecell}

\newcommand{\tsun}{T_{\odot}}
\newcommand{\kms}{\,\mathrm{km}\,\mathrm{s}^{-1}}

\newcommand{\pb}{P_{\text{b}}}

\newcommand{\ntoa}{1399}

\newcommand{\fhour}{^{\mathrm{h}}}
\newcommand{\fmin}{^{\mathrm{m}}}
\newcommand{\fsec}{\mbox{\ensuremath{.\!\!^{\mathrm{s}}}}}
\newcommand{\fdeg}{^{\circ}}
\newcommand{\fdegdecimal}{\mbox{\ensuremath{.\!\!^{\circ}}}}

\newcommand{\mone}{$1.306$}
\newcommand{\mtwo}{$1.299$}

\newcommand{\vt}{$43^{+51}_{-34}$}
\newcommand{\vtot}{$49^{+77}_{-30}$}

\newcommand{\psr}{PSR~J1829$+$2456}

\newcommand{\msun}{\,M_{\odot}}
\newcommand{\degrees}{^\circ}
\newcommand{\omdot}{\dot{\omega}}

\title[\psr]{Precise mass measurements for the double neutron star system J1829$+$2456}

\author[Haniewicz et~al.]{H.~T.~Haniewicz$^{1}$\thanks{E-mail: h.haniewicz@uea.ac.uk},
R.~D.~Ferdman$^{1}$,
P.~C.~C.~Freire$^{2}$,
D.~J.~Champion$^{2}$,
\newauthor K.~A.~Bunting$^{3}$,
D.~R.~Lorimer$^{4}$$^,$$^{5}$
and M.~A.~McLaughlin$^{4}$$^,$$^{5}$ \vspace{0.5cm} \\
$^{1}$Faculty of Science, University of East Anglia, Norwich Research Park, Norwich NR4 7TJ, UK \\
$^{2}$Max-Planck-Institut f{\"u}r Radioastronomie, Auf dem H{\"u}gel 69, D-53121 Bonn, Germany \\
$^{3}$School of Science and Technology, Nottingham Trent University, College Drive, Clifton, Nottingham NG11 8NS, UK \\
$^{4}$Department of Physics and Astronomy, West Virginia University, P.O. Box 6315, Morgantown, WV 26506, USA \\
$^{5}$Center for Gravitational Waves and Cosmology, West Virginia University, Chestnut Ridge Research Building, Morgantown, \\ WV 26505, USA
}

\date{Accepted 2020 November 3. Received 2020 October 7; in original form 2020 July 14}

\pubyear{2020}

\begin{document}
\label{firstpage}
\pagerange{\pageref{firstpage}--\pageref{lastpage}}
\maketitle

% Abstract of the paper
\begin{abstract}
\psr{} is a radio pulsar in a relativistic binary system with another neutron star. It has a rotational period of 41 ms and a mildly eccentric ($e = 0.14$) 28-hr orbit. We have continued its observations with the Arecibo radio telescope and have now measured the individual neutron star masses of this system: The pulsar and companion masses are $\mone\,\pm\,0.007\msun$ and $\mtwo\,\pm\,0.007\msun$ ($2\sigma - 95\%$ confidence, unless stated otherwise), respectively. We have also measured the proper motion for this system and used it to estimate a space velocity of \vtot $\kms$ with respect to the local standard of rest. The relatively low values for companion mass, space velocity and orbital eccentricity in this system make it similar to other double neutron star systems in which the second-formed neutron star is thought to have formed in a low-kick, low mass-loss, symmetric supernova.
\end{abstract}

% Select between one and six entries from the list of approved keywords.
% Don't make up new ones.

%%%%%%%%%%%%%%%%%%%%%%%%%%%%%%%%%%%%%%%%%%%%%%%%%%

%%%%%%%%%%%%%%%%% BODY OF PAPER %%%%%%%%%%%%%%%%%%

\section{Introduction}

The observational study of neutron stars (NSs) in binary systems began with the discovery of the relativistic binary pulsar PSR~B1913$+$16 by \citet{Hulse1975}, who recognised its value for precise tests of the predictions of general relativity (GR) in the strong-field regime \citep{DDGR1989}, as well as tests of alternative theories of gravity (see e.g., the scalar-tensor theories of \citealt{Damour1992, Damour1993, Damour1996}). Notable examples include the aforementioned PSR~B1913$+$16, observations of which have shown agreement with GR predictions of orbital decay rate due to gravitational wave emission \citep{Weisberg2016}; PSR~B1534$+$12 \citep{Stairs2002, B1534}; the double pulsar (PSR~J0737$-$3039A/B), which has given the most stringent test of GR in the strong-field regime \citep{Kramer2006}; PSR~J1738+0333, which has introduced stringent constraints on the nature of gravitational radiation and on the aforementioned scalar-tensor theory \citep{Freire2012}; PSR~J0337$+$1715, a pulsar in a triple system that has provided the most constraining limits to date on violation of the universality of free fall and has also provided very stringent limits on scalar-tensor theory and other alternative theories of gravity \citep{Archibald2018, Voisin2020}; and PSR~J1141$-$6545, whose white dwarf companion is observed to undergo relativistic frame-dragging during its rotation \citep{Krishnan2020}.

Observations of pulsars in binaries also allow us to probe binary formation and evolution. In particular, studies of double neutron star (DNS) systems can provide insight into the formation of the second-formed NS and ultimately the physics behind the second supernova, of which the DNS is a remnant (e.g., \citealt{Tauris2017}). The progenitor binary systems of most DNS systems contain stars of mass $\gtrsim 8\msun$. The more massive star is first to end its main-sequence, the remnant of which is a fast spinning neutron star and a main sequence binary companion.

Amongst the known Galactic DNS population, it is becoming clear that there are two principal post--second supernova evolutionary channels, observationally distinguished by their companion masses, orbital eccentricities and space velocities. Systems with a high eccentricity, high companion mass and high peculiar space velocity ($v^{\text{LSR}}$) compared to the overall DNS population, such as PSRs B1534+12 and B1913+16 indicate that they are the result of a high mass-loss, asymmetric second--SN from a massive progenitor that led to a large natal kick from the system \citep[e.g.][]{Wex2000}.

In contrast, systems with a low eccentricity (when compared to other DNSs), low companion mass and small $v^{\text{LSR}}$, such as PSRs J0737$-$3039 \citep{Kramer2006, Ferdman2013}, J1756$-$2251 \citep{Ferdman2014}, J1913+1102 \citep{Ferdman2020} and J1946$+$2052 \citep{Stovall2018} are theorized to have all undergone a different evolutionary track from the aforementioned binaries after the first supernova. As with most DNS systems, it is thought that the Roche lobe overflow (RLO) of the companion star, as it evolves off the main sequence, results in a common envelope (CE) in which the first-formed NS is embedded.  After inspiral of the NS due to dynamical friction and the ultimate ejection of the common envelope, a NS-helium (He) star binary is left behind \citep{Tauris2017}. Depending on the mass of the He-star and the orbital separation, the surface layers of the companion may be tidally stripped following Case BB RLO mass accretion to the NS, causing it to gain angular momentum and spin up to have a rotation period of tens of milliseconds \citep{Tauris2015}. If the He-star is massive enough, it also undergoes a supernova via a rapid electron capture (EC) to an iron or possibly an O--Ne--Mg core \citep{Tauris2017} or ultra-stripped iron core-collapse (FeCCSNe) \citep{Tauris2013}. The time scales for EC and FeCCSNe are much faster than the timescales for non-radial hydrodynamical instabilities to occur \citep{Nomoto1984, Zha2019}, leading to a low-kick symmetric supernova resulting in a $\lesssim \, 0.1$ eccentricity increase \citep{Tauris2017}. A system that has undergone this process is expected to have a correspondingly low observed eccentricity and space velocity ($\lesssim 100$ km s$^{-1}$).

If these scenarios are correct, there should be a correlation between NS mass and inferred SN kick velocity. This seems to be the case \citep{Tauris2017}; however the number of DNSs with good mass and proper motion measurements is still less than half of the known sample of 19 DNSs in the Galactic disk, which results in low-number statistics. It is therefore imperative not only that we discover, but also that we measure masses, proper motions and other parameters for as many DNS systems as possible in order to expand on this relatively small population. This has additional benefits: recently, two DNSs with asymmetric NS masses have been discovered \citep{Martinez2015, Ferdman2020}. This has not only expanded the range of NS masses observed in DNSs, but showed the existence of asymmetric systems. The latter system is expected to merge within 470 Myr; this suggests that the population of such merging asymmetric DNSs might be substantial, with a fraction of about 10\% of the known merging DNS population; however, our knowledge is still limited by the small numbers of the DNS with well-known masses. Establishing more firmly the size of this population will be of particular importance for the interpretation of DNS mergers in LIGO/Virgo/Kagra data.

\psr{} is a recycled pulsar with a rotational period of 41 ms, and is a member of a DNS system in a 28-h (1.18-d), mildly eccentric  orbit ($e = 0.14$). It was initially discovered and timed by \citet{Champion2004} from data taken during a 1999 drift-scan survey using the 430-MHz Gregorian dome receiver system at the Arecibo radio telescope. At the time of its discovery, the dispersion measure (DM) was found to be 13.9\,pc\,cm$^{-3}$, which implied a distance of $1.2 \pm 0.36$ kpc to the pulsar, estimated from the NE2001 Galactic ionized electron distribution model \citep{Cordes2003}. However, due to the existence of the Gould Belt, a dense region of gas and young stellar populations along the line of sight to \psr{} \citep{Gehrels2000, Grenier2000}, this distance has likely been overestimated. A more reliable estimated distance may come from using the YMW16 electron distribution model \citep{Yao2017}, which includes several local features such as those due to the Local Bubble, and adds a fourth spiral arm to the model of the Milky Way. YMW16 models the distance of PSR~J1829$+$2456 to be $0.91 \pm 0.18$ kpc.

Soon after its discovery, the advance of periastron ($\omdot$) of \psr\ was found to be $0.2919\,\pm 0.0016\,\degrees\,\text{yr}^{-1}$, leading to a total mass ($M_{\text{tot}}$) estimate of $2.59\,\pm 0.02\msun$ \citep{Champion2005}. However, only limiting values of the pulsar and companion mass could be found, with $m_{\text{p}} < 1.38\msun$ and $1.22\msun < m_{\text{c}} < 1.38\msun$. Although these mass limits alone do not conclusively determine the companion to be a NS (as opposed to a massive white dwarf), the moderate eccentricity of the orbit in tandem with these mass limits, as well as the spin period of tens of milliseconds and a small $\dot{P}$, which give a large characteristic age of 13 Gyr and a small surface magnetic field strength of $1.4 \times 10^{9}$ G (characteristics generally observed post-recycling), implied that the system is likely to be a DNS.
Although the recycling of the first-formed pulsars in these systems likely circularized the orbits, as observed for high-mass X-ray binaries, DNS systems are expected to have at least $10^3$ times higher eccentricities than NS$-$WD systems with the same orbital periods due to large, near-instant mass loss that occurs during the supernova that forms the second NS and its associated kick. By contrast, in NS-WD systems where the NS is recycled, the orbit retains the low eccentricity associated with their X-ray binary phase, since no second supernova disrupts the system \citep[see e.g.][]{Antoniadis2013, Wang2017}.

The new observations for \psr{} were predicted to allow us to significantly determine the system component masses as well as better constrain the proper motion. This would allow for tighter constraints on binary evolution models for DNS systems and determine this system's evolutionary track in the context of the wider DNS population.

\section{Observations and timing analysis}
\label{sec:obs}

Initial observations of \psr\ began in May 2003 (MJD 52785) with the Arecibo telescope, using the Penn State Pulsar Machine (PSPM) at a centre frequency of 430\,MHz, and the Wideband Arecibo Pulsar Processor (WAPP) centred at 1400\,MHz. Several observations were carried out using the Green Bank telescope (GBT) at 350\,MHz in August 2006, only 10 pulse time-of-arrival (TOA) measurements could be salvaged due to overly pervasive radio frequency interference (RFI) in that data set. A full description of the data set and its analysis can be found in \citet{Champion2004} and \citet{Champion2005}.

The most recent observing campaigns for \psr\ have been running since July 2017 (MJD 57950), and this work analyzes data taken until June 2020 (MJD 59014). All these observations were conducted at the Arecibo radio telescope roughly every 4 weeks using the Puerto Rico Ultimate Pulsar Processing Instrument (PUPPI) coherent de-dispersion backend. Two frequency bands were used during these observations with centre frequencies of $\sim 1400$\,MHz and $\sim 430$\,MHz, over bandwidths of 800\,MHz and 100\,MHz, respectively. In the second year of PUPPI observations, we conducted a dense campaign of four epochs within one week in order to provide better orbital phase sampling.

A standard profile for each band was created in an iterative manner. We began by averaging all the folded data from a particular backend and using the resulting profile's total intensity as the template. RFI excision was conducted on data from the correct frequency band by cross-referencing against this profile. This was done by fitting the current template to four Gaussian curves to obtain a smooth standard profile, allowing for a clear distinction between on-pulse and off-pulse regions. Individual profiles were then rejected if their off-pulse RMS was a 95\% outlier to the overall off-pulse RMS distribution. After RFI excision, a new profile was constructed in the same way as described above by averaging the newly RFI excised data.

The cleaned data were flux calibrated by comparing against observations\footnote{Data for this continuum source was provided by the NANOGrav collaboration.} of the stably polarised quasar QSO~B1442 (J1445+0958) as a continuum source at the closest available dates to the PSR~J1829+2456 observations; the largest time difference between our data and corresponding calibration data sets was nine days. After flux calibration, the fully processed data were once again used to create a standard profile for the band. Following this, an initial set of pulse times of arrival (TOAs) were generated, and a timing solution was fit to these TOAs. This was used to re-fold and phase-realign the PUPPI data.

\begin{figure}
    \includegraphics[width=\columnwidth]{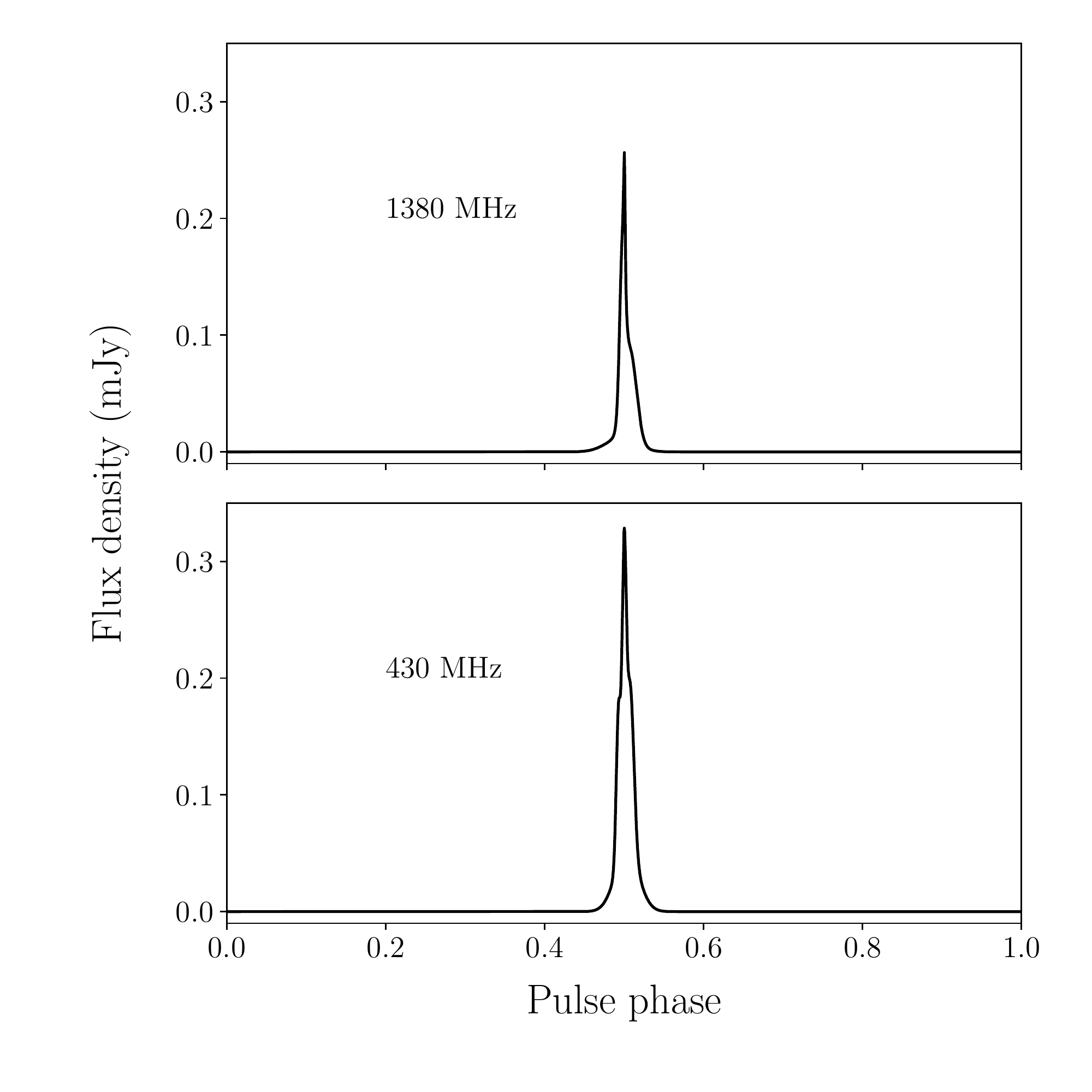}
    \caption{The noise-free templates of \psr, ($T_{\nu}$ in the text) constructed from all data taken with the PUPPI coherent dedispersion backend at Arecibo, up to and including MJD 58948. Top: L-band template. Bottom: 430-MHz template. Both templates were fit with four Gaussian functions and the visible peaks are to scale relative to each other.}
    \label{fig:lbandtemp}
\end{figure}

Using the same RFI excision masks and calibration factors as calculated in the previous step, the phase-aligned profiles were used to create the final standard profiles for each observing band, $T_{\nu}$ (where $\nu$ is the band centre frequency), by fitting the resulting points to four Gaussians. This final step was done in order to ensure the template accurately reflected the intrinsic pulse shape, resulting in minimised timing residual errors. The final standard profiles for both the 430\,MHz and L-band data are shown in Figure \ref{fig:lbandtemp}. All data manipulation was administered using the PSRVoid and PyPulse Python packages\footnote{[ascl:2007.007], [ascl:1706.011]}.

% Template figure

In all we calculated 1246 new pulse times-of-arrival from the data set by determining a phase offset for each resulting data profile through cross-correlation with the standard profile for each observing band. This phase shift was then converted to a time offset using the rotation period at the epoch corresponding to the individual data profile \citep{Taylor1992}. TOAs were created from time-averaged sub-integrations of approximately 9 minutes for the 430 MHz data, corresponding to 10 TOAs per observation, and about 3 minutes for the L-band data. The L-band data were divided into four frequency subbands centred at approximately 1680\,MHz, 1480\,MHz, 1280\,MHz and 1080\,MHz. Due to receiver cut-off near the lower L-band frequencies, there is very little visible profile below 1100\,MHz and we therefore omit it from the analysis, as well as TOAs with large uncertainties ($\gtrsim 25\,\mu$s). In total, this resulted in 934 L-band and 314 430\,MHz TOAs.

The TOAs were then appended to the 153 existing TOAs from older observing campaigns for this pulsar. These were fit within the \textsc{tempo2} pulsar timing software package \citep{Hobbs2006, Edwards2006} using the JPL DE435 Solar System emphemeris model \citep{DE435} and the TT(BIPM19) \citep{Guinot1988} clock correction in order to convert the observatory time-stamp assigned to each profile to the GPS time standard. These were ultimately, to the Solar System barycentre (SSB) which is, to good approximation, an inertial reference frame. Where TT(BIPM19) could not be used (i.e. for the final three days of data), a correction was made in accordance with BIPM guidelines\footnote{ftp://ftp2.bipm.org/pub/tai/ttbipm/TTBIPM.2019}. \textsc{tempo2} fits all TOAs to an existing model ephemeris via a weighted least-squares fit, and outputs a set of timing residuals, which are the differences between the observed TOAs and those predicted from the current model. In all, \ntoa\ TOAs were fit, spanning 17.1 years in total, at frequencies centred around 350\,MHz, 430\,MHz and the three remaining L--band subbands mentioned above.

Residual errors were calculated from the uncertainty in the phase shift calculated in the cross-correlation process.
A breakdown of each observation campaign is shown in Table \ref{tab:observations}. Due to the gap in time between the PSPM/WAPP and the PUPPI data sets, it is not expected that the latter is fully phase connected with the former. To account for this, along with any other backend-specific systematics, phase offsets were fit between each set of TOAs obtained from each different backend.

%%%%%%%%%%%%%%%%%%%%%%%%%%%%%%%%%%%%%%%%%%%%%%%%%%

%%%%%%%%%%%%%%%%%%%% OBSERVATIONS TABLE %%%%%%%%%%%%%%%%%%

\begin{table*}
	\centering
	\caption{Summary of time-of-arrival data for \psr{}. \label{tab:observations}}
	\begin{tabular}{llccrccc}
		\hline
Telescope & Instrument & \makecell{Centre  Frequency \\ (MHz)} & \makecell{Bandwidth \\ (MHz)} & \makecell{Span \\ (MJD)} & \#TOAs & Weighted $\chi^{2\dagger}$ & \makecell{Weighted \\ RMS$^{\dagger}$ ($\mu$s)} \\
\hline
Arecibo & PSPM & 434.0 & 7.68 & 52785 $-$ 53905 & 117 & 1.1873 & 17.0062 \\
 & PSPM & 331.0 & 7.68 & 53027 $-$ 53476 & 2 & 1.7937 & 19.9603 \\
 & WAPP & 1378.6 & 100 & 54588 $-$ 54835 & 11 & 2.0001 & 5.7050 \\
 & WAPP & 319.6 & 100 & 54647 $-$ 54835 & 11 & 0.9655 & 9.6925 \\
 & PUPPI & 1384.4 & 800 & 57950 $-$ 58948 & 934 & 1.1726 & 3.4843 \\
 & PUPPI & 427.2 & 100 & 57950 $-$ 59014 & 314 & 0.9938 & 10.7057 \\
GBT & GASP & 350.0 & 16 & 52972 $-$ 52973 & 10 & 0.5231 & 32.5539 \\
\hline
\multicolumn{2}{l}{Overall} & 676.5 & -- & 52785 $-$ 59014 & \ntoa & 1.1584 & 3.9640 \\
\end{tabular}
\flushleft
$^{\dagger}$From the DDGR binary fit.
\end{table*}

%%%%%%%%%%%%%%%%%%%% PARAMETER TABLE %%%%%%%%%%%%%%%%%%
% Residual figure
\begin{figure*}
 \centering
 \includegraphics[width=0.8\textwidth]{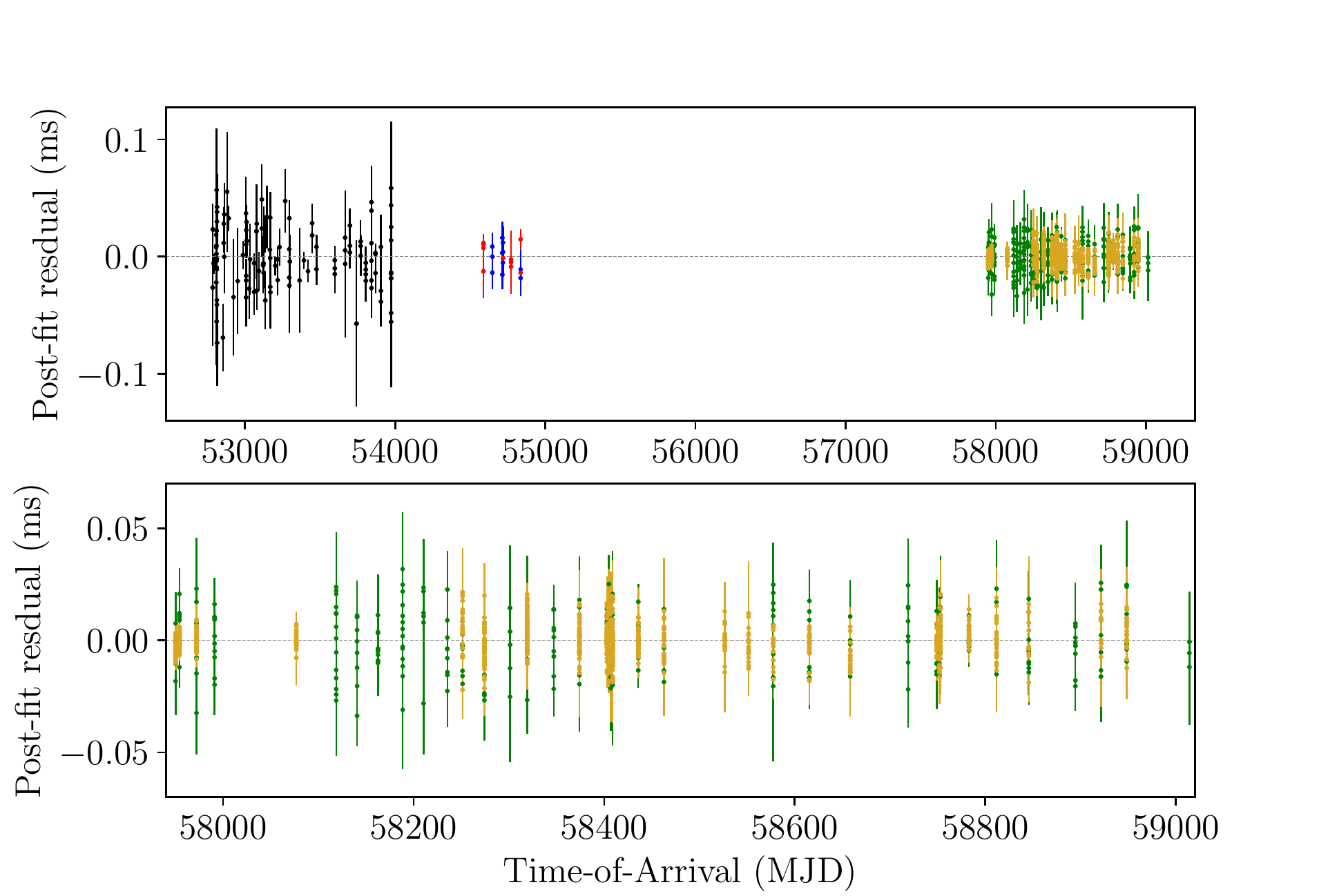}
 \caption{Post-fit residuals in milliseconds, as a function of TOA in MJD, for \psr\ determined by the DDGR timing model. Top: all available Times-of-Arrival. Bottom: the new observations in green (430\,MHz) and gold (1400\,MHz).}
 \label{fig:resids}
\end{figure*}

\begin{table*}
	\centering
	\caption{Timing solution for \psr{}. \label{tab:1829_params}}
	\begin{tabular}{lcc}
		\hline
\multicolumn{3}{c}{Fit and data-set} \\
\hline
Data span (yr)\dotfill & \multicolumn{2}{c}{17.1} \\
Date range (MJD)\dotfill & \multicolumn{2}{c}{$52785.3 - 59015.3$} \\
Number of TOAs\dotfill & \multicolumn{2}{c}{\ntoa} \\
Solar System ephemeris\dotfill & \multicolumn{2}{c}{DE435} \\
Clock correction procedure\dotfill & \multicolumn{2}{c}{TT(BIPM19)} \\
Reference timing epoch (MJD)\dotfill & \multicolumn{2}{c}{$55899.8$} \\
Binary model \dotfill & DDH & DDGR \\
RMS timing residual ($\mu$s)\dotfill & 3.967 & 3.964 \\
\hline
\multicolumn{3}{c}{Observed quantities} \\
\hline
Right ascension, $\alpha_{\text{J2000}}$\dotfill & \multicolumn{2}{c}{$18\fhour29\fmin34\fsec66838(6)$} \\
Declination, $\delta_{\text{J2000}}$\dotfill & \multicolumn{2}{c}{$24\fdeg56\farcm18\farcs2007(12)$} \\
Rotation frequency, $\nu$ (s$^{-1}$)\dotfill & $24.384401411044(6)$ & $24.384401411040(6)$ \\
First derivative of rotation frequency, $\dot{\nu}$ (s$^{-2}$)\dotfill & $-2.9403(13) \times 10^{-17}$ & $ -2.9395(14) \times 10^{-17}$ \\
Dispersion measure, DM (cm$^{-3}$pc)\dotfill & $13.706(2)$ & $13.707(2)$ \\
$\dot{\text{DM}}$ (cm$^{-3}$ pc\ yr$^{-1}$)\dotfill & $-0.0019(4)$ & $-0.0022(4)$ \\
$\ddot{\text{DM}}$ (cm$^{-3}$ pc\ yr$^{-2}$)\dotfill & $0.00023(4)$ & $0.00024(4)$ \\
$\dddot{\text{DM}}$ (cm$^{-3}$ pc\ yr$^{-3}$)\dotfill & $-7.0(1.6) \times 10^{-6}$ & $-7.1(1.6) \times 10^{-6}$ \\
Proper motion in right ascension, $\mu_{\alpha}$ (mas\,yr$^{-1}$)\dotfill & $-5.51(5)$ & $-5.51(6)$ \\
Proper motion in declination, $\mu_{\delta}$ (mas\,yr$^{-1}$)\dotfill & $-7.75(7)$ & $-7.82(8)$ \\
Binary period, $\pb$ (d)\dotfill & $1.176027952868(11)$ & $1.17602795281(15)$ \\
Orbital eccentricity, $e$\dotfill & $0.13914374(13)$ & $0.13914387(11)$ \\
Projected semi-major axis of orbit, $x$ (lt-s)\dotfill & $7.236845(2)$ & $7.236844(5)$ \\
$\dot{x}$ (lt-s s$^{-1}$)\dotfill & \multicolumn{2}{c}{$-2.3(5) \times 10^{-14}$} \\
Longitude of periastron, $\omega$ ($\degrees$)\dotfill & \multicolumn{2}{c}{$229.9353(2)$} \\
Epoch of periastron, $T_0$ (MJD)\dotfill & \multicolumn{2}{c}{$52848.5797762(7)$} \\
Advance of periastron, $\dot{\omega}$ ($\degrees\,$yr$^{-1}$)\dotfill & $0.293189(14)$ & -- \\
Orbital period decay, $\dot{P}_{\text{b}}$ \dotfill & $-2.9(1.2) \times 10^{-14}$ & -- \\
Non--GR contribution to orbital decay, $\dot{P}_{\text{b}}^{\text{X}}$ \dotfill & -- & $-2.3(1.1) \times 10^{-14}$ \\
Ratio of Shapiro harmonics, $\varsigma$ \dotfill & $0.778(2)$ &  \\
Companion mass, $m_{\text{c}}$ ($\msun$)\dotfill & -- & $1.299(4)$ \\
Total system mass, $M_{\text{tot}}$ ($\msun$)\dotfill & -- & $2.60551(19)$ \\
\hline
\multicolumn{3}{c}{Derived quantities} \\
\hline
Rotation period, $P$, (ms)\dotfill & $41.009823581203(11)$ & $41.009823581195(11)$ \\
First derivative of rotation period, $\dot{P}$\dotfill & $4.945(2) \times 10^{-20}$ &  $4.944(2) \times 10^{-20}$ \\
Intrinsic spin-down rate, $\dot{P}_{\text{int}}$ \dotfill & $4.36(9) \times 10^{-20}$ & $4.35(9) \times 10^{-20}$ \\
Galactic longitude, $\ell$\dotfill & \multicolumn{2}{c}{$53\fdegdecimal3426(11)$} \\
Galactic latitude, $b$\dotfill & \multicolumn{2}{c}{$15\fdegdecimal6119(12)$} \\
NE2001 DM-derived distance (kpc)\dotfill & \multicolumn{2}{c}{$1.20(36)$} \\
YMW16 DM-derived distance (kpc)\dotfill & \multicolumn{2}{c}{$0.91(18)$} \\
Height above Galactic plane, $z$ (kpc)\dotfill & \multicolumn{2}{c}{0.24(5)} \\
Total proper motion, $\mu_{\text{tot}}$ (mas\,yr$^{-1}$)\dotfill & $9.52(7)$ & $9.56(7)$ \\
Transverse velocity, $v_{\text{trans}}$ ($\kms$)\dotfill & \multicolumn{2}{c}{\vt} \\
Total peculiar velocity, $v_{\text{tot}}$ ($\kms$)\dotfill & \multicolumn{2}{c}{\vtot} \\
Characteristic age, $\tau_{\text{c}}$ (Gyr)\dotfill & \multicolumn{2}{c}{$13$} \\
Surface magnetic field strength, $B_{\text{s}}$ ($10^9\,$G)\dotfill & \multicolumn{2}{c}{1.44} \\
Mass function, $f$ ($\msun$)\dotfill & $0.2942356(3)$ & $0.2942355(5)$ \\
Einstein delay, $\gamma$ (s)\dotfill & -- & $0.00143977$ \\
Inclination of orbit, $i$ ($\degrees$)\dotfill & -- & $75.8(7)^{*}$ \\
Orthometric amplitude of Shapiro delay, $h_3$ ($\mu$s)\dotfill & -- & $3.02$ \\
Pulsar mass, $m_{\text{p}}$ ($\msun$)\dotfill & -- & $1.306(4)^{\dagger}$ \\
\end{tabular}
\flushleft
PK parameters $\omdot$, $\dot{P}_{\text{b}}$ and $\varsigma$ were measured using the orthometric parameterised Shapiro delay Damour-Deruelle (\textsc{DDH}) timing model \citep{DDH} in \textsc{tempo2} whereas the quoted masses, $\gamma$ and the inclination angle were measured and derived assuming GR as the correct theory of gravity (DDGR) \citep{DD1986}. Figures in parentheses represent the nominal $1\sigma$ (68\%) uncertainties in the least-significant digits quoted. Time offsets between telescopes and different instruments were also fit for using Arecibo's PSPM backend at 430\,MHz as a reference, however they are not astrophysical, so they are not shown here. Using any other backend as the basis for our jumps gave consistent results. \\
$^{*}$Calculated using the binary mass function and the component masses in the relation $f = (m_{\text{c}} \sin{i})^3 / M_{\text{tot}}^2$. The reported uncertainty is a result of error propagation on the masses and mass function. \\
$^{\dagger}$Derived from $M_{\rm tot} - m_{\text{c}}$.
\end{table*}

\subsection{Binary models}
\label{sec:models}

The binary motion can be described, to first approximation, by five Keplerian orbital elements: orbital period ($\pb$); projection of the semi-major axis onto the line of sight ($x \equiv a \sin{i}$), where $a$ is the semi-major axis of the pulsar's orbit and $i$ is the orbital inclination; orbital eccentricity ($e$); longitude of periastron ($\omega$); and epoch of periastron passage ($T_0$). We also fit for any significant relativistic perturbations to the Keplerian motion. Two binary-timing models were used, both based on the Damour-Deruelle (DD) timing model \citep{DD1986}. The first of these is the DDGR model, which considers general relativity to be the correct theory of gravity, and where we fit only for the Keplerian parameters, the total mass and the companion mass to describe the timing of the pulsar.

The second is the DDH model \citep{DDH}. Like the DD model, this parameterises the relativistic perturbations in the timing using the so-called ``post-Keplerian'' (PK)  parameters. This is done in a theory-independent way, so that the
description of the motion can be interpreted by a wide class of theories of gravity \citep{DT92}.
The DDH model reparameterises the traditional Shapiro delay ``range'' ($r$) and ``shape' ($s$) parameters (describing the pulse arrival delay due to the gravitational field of the companion star when the pulsar is at superior conjunction) in terms of the so-called orthometric parameters $h_3$ (the orthometric amplitude) and $\varsigma$ (the orthometric ratio).
The advantage of the DDH model is that the covariances between its Shapiro delay parameters is generally much lower than between $r$ and $s$ in the DD model for larger orbital inclinations.

Under specific theories of gravity, the PK parameters are related to the masses and orbital inclination; in the case of GR, we have \citep{DD1985, DD1986, DDH}:

\begin{align}
\omdot & = 3\tsun^{2/3} \left( \frac{\pb}{2\pi} \right)^{-5/3} \frac{(m_{\text{p}} + m_{\text{c}})^{2/3}}{1 - e^2} \label{eq:PKomdot} \\
\gamma & = \tsun^{2/3} \left( \frac{\pb}{2\pi} \right)^{1/3} e\frac{m_{\text{c}}(m_{\text{p}} + 2m_{\text{c}})}{(m_{\text{p}} + m_{\text{c}})^{4/3}}\label{eq:PKgamma} \\
\dot{P}_{\text{b}} & = -\frac{192\pi}{5} \tsun^{5/3} \left( \frac{\pb}{2\pi} \right)^{-5/3} \frac{m_{\text{p}} m_{\text{c}} \left( 1 + \frac{73}{24}e^2 + \frac{37}{96}e^4 \right)}{(m_{\text{p}} + m_{\text{c}})^{1/3} \left( 1 - e^2 \right)^{7/2} } \label{eq:PKpbdot} \\
\varsigma & = \frac{\sin i}{1 + \sqrt{1 - \sin i^2}} \label{eq:PKs} \\
h_3 & = \tsun m_{\text{c}} \varsigma^3\label{eq:PKr}
\end{align}
where $\omdot$ is periastron advance, $\gamma$ is the Einstein delay parameter, $\dot{P}_{\text{b}}$ is the orbital decay, $h_3$ and $\varsigma$ are the orthometric Shapiro parameters, and $\tsun \equiv G M_{\odot} / c^3 = 4.9254909476412675\ \mu$s. Therefore, two PK parameter measurements will allow us to solve for each component mass, and each additional PK parameter measurement provides a unique check for consistency, and therefore, a test of GR.

% Contour figure
\begin{figure}
 \centering
 \includegraphics[width=\columnwidth]{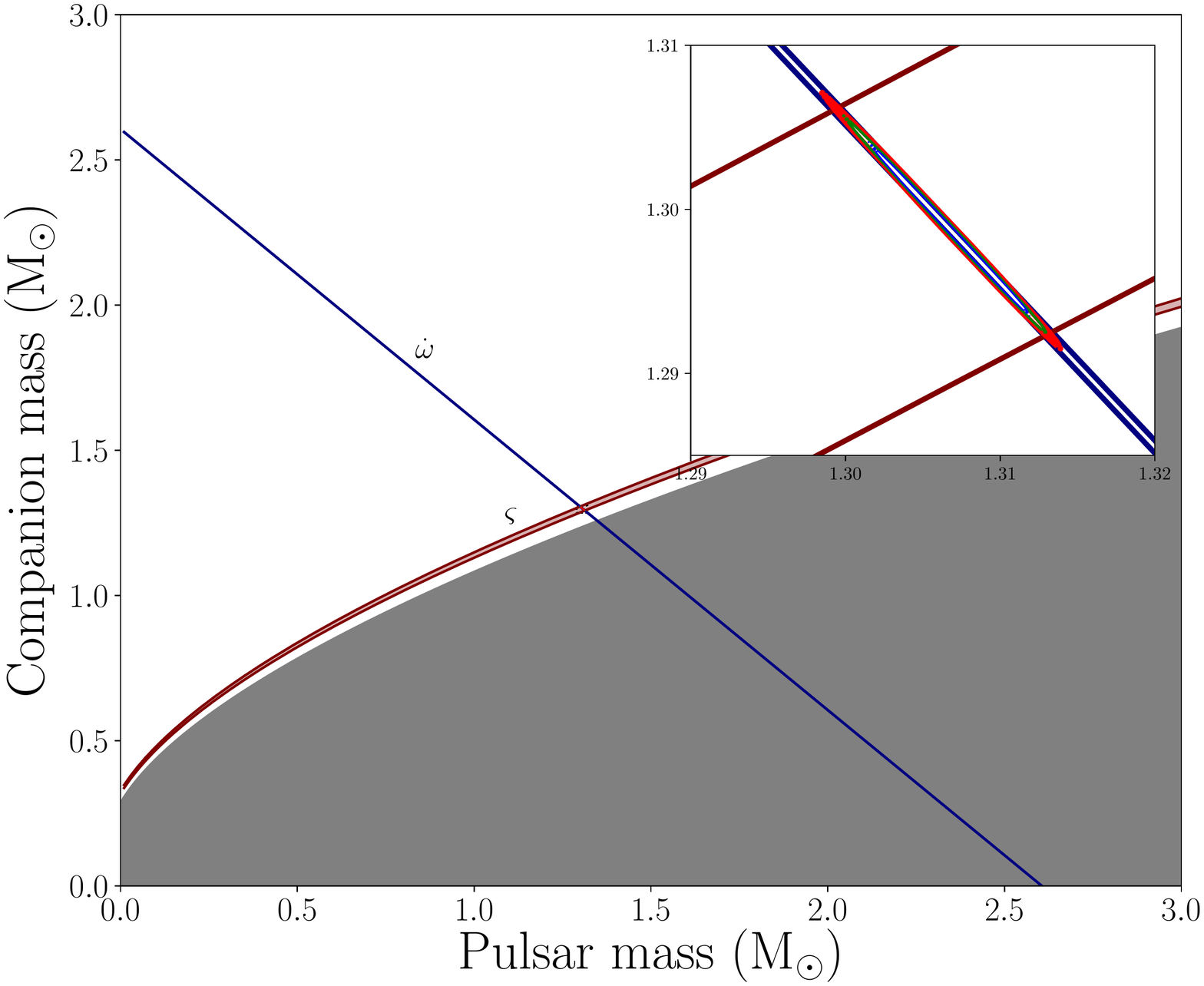}
 \caption{Main window: Mass-mass diagram for \psr{} showing the GR-derived mass constraints from the PK measurements determined by DDH, as reported by \textsc{tempo2}. The dark blue region is $\omdot$ and the maroon region is $\varsigma$. The grey shaded region represents $\sin i > 1$. Inset: The green contoured region represents the 95\% confidence region for the pulsar and and companion masses based on the DDGR model, which assumes general relativity for the timing fit.}
 \label{fig:DDcontours}
\end{figure}

Figure \ref{fig:resids} shows the timing residuals from the DDGR model fit and Table \ref{tab:1829_params} gives the resulting post-fit timing parameters from \textsc{tempo2} using both the DDH and DDGR models.

\section{Results and discussion}

Newly measured properties of \psr\ have shown the system to be similar to several other DNS systems for which component masses have been measured. Given the evolutionary relationship between the masses and the parameters discussed, it is believed that these systems underwent similar evolutionary processes \citep{Tauris2017}. Table~\ref{tab:PSRs} compares the known recycled DNS systems for which mass measurements have been made or bounded.

\subsection{Component masses} \label{sec:masses}

From the DDGR model, we measure a total system mass of $2.60551(38) \msun$, which is fully consistent with the latest value from \citet{Champion2005}. The component masses are $\mone\,\pm\,0.007\msun$ for the pulsar and $\mtwo\,\pm\,0.007\msun$ for the companion; unexceptional among the DNS population (where precise mass measurements are known). In order to verify these uncertainties, we used a Bayesian method, a contour grid was sampled over pulsar and companion mass with confidence levels being obtained by calculating the likelihood based on the $\chi^2$ of the fit at each grid point (Figure \ref{fig:DDcontours} inset).

The individual constraints as found with DDGR require the presence of a detectable Shapiro delay in addition to the rate of advance of periastron, otherwise the masses would be unbounded in GR. This is because our current estimate of $\gamma$ is not sufficiently accurate.
It has been shown that, for sufficiently wide orbits, the derivative of the projected semi-major axis can be highly covariant with $\gamma$, which itself is covariant with the current value for $x$ as well as the proper motion \citep[][equations 25 and 43 respectively]{Ridolfi2019}. We have calculated the absolute maximum contribution to $\dot{x}$ due to proper motion to be $2.6 \times 10^{-15}$ lt-s s$^{-1}$. However, this estimate is one order of magnitude smaller than the value for $\dot{x}$ as given by the DDGR fit, which itself is $4\,\sigma$ significant. The contribution to $\dot{x}$ due to the Lense-Thirring effect \citep{Krishnan2020} was also found to be insignificant when compared with the uncertainty in $\dot{x}$. Because of this unexplained value of $\dot{x}$, we cannot reliably measure $\gamma$.
Given the masses determined by the DDGR model, this is predicted by GR to have a value of 1.44\,ms (Equation \ref{eq:PKgamma}).

Since we believe that a Shapiro delay signal must be present, we used the DDH model to measure it. In doing so, we held fixed $\gamma$ as well as $h_3$ at their GR-derived values (equations \ref{eq:PKs} and \ref{eq:PKr}), with all other model parameters allowed to vary,
including $\dot{\omega}$ and $\varsigma$.
The general relativistic mass constraints introduced by the resulting values of the latter PK parameters can be compared with the DDGR-derived mass contours; this is shown in Figure \ref{fig:DDcontours}. The good agreement with the DDGR contours demonstrates that these are the two effects that allow us to measure the component masses for \psr{}.

\subsection{Space velocities} \label{sec:velocities}

\begin{figure}
 \centering
 \includegraphics[width=\columnwidth]{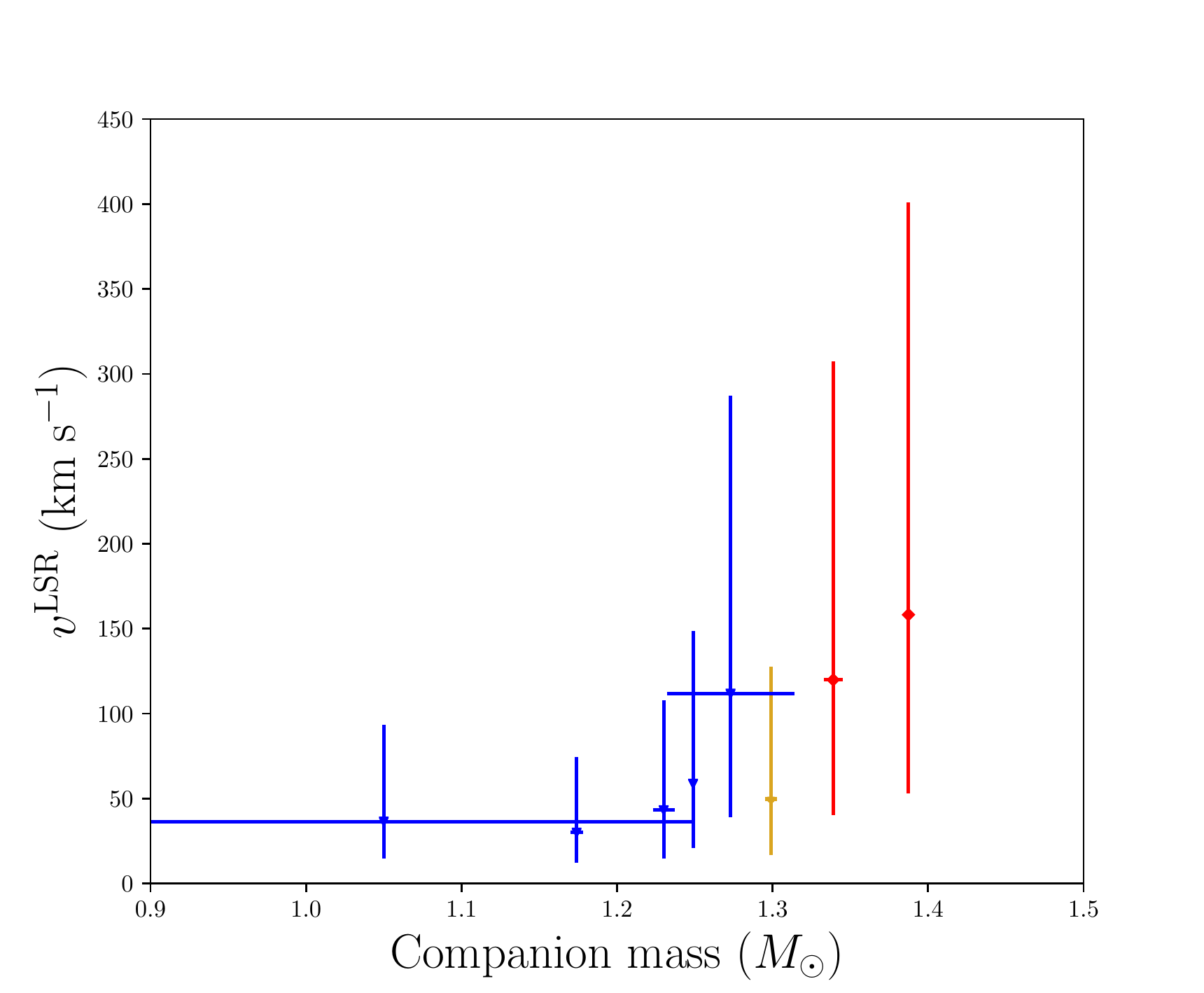}
 \caption{The YMW16$-$derived space velocities, taken with respect to the Local Standard of Rest (LSR), with respect to the mass of the second-formed (companion) NS, of all DNSs with known masses and component proper motions. The gold star point represents \psr{}. DNSs thought to have followed a similar evolution to \psr{} are represented by blue triangles. Errors in $v^{\text{LSR}}$ and $m_{\text{c}}$ are the $2\sigma$ confidence level values.}
 \label{fig:velocity}
\end{figure}

As a result of the newly analysed campaign, the component proper motions for \psr\ have been precisely determined. In order to compare this with other DNS systems, the three-dimensional space velocities for all binary systems containing pulsars with known individual proper motions were calculated. This involves a two-prior Monte Carlo approach to estimate both the distance ($d$, using YMW16 estimates) and the transverse velocity ($v_{\text{trans}}$), each sampled from a Gaussian distribution with the $1\,\sigma$ width equal to their uncertainties. The error on $v_{\text{trans}}$ was propagated through the error on $d$ and the uncertainty in the proper motion. Since the radial velocity ($v_{\text{r}}$) is not possible to determine using pulsar timing, we have randomly sampled $v_{\text{r}}$ from a distribution which is uniform in $\cos i$, and calculated the total space velocity as follows \citep{Tauris2017}:

\begin{align}
    v_{\text{tot}}\left( d, i \right) &= \sqrt{v^2_{\text{r}} + v^2_{\text{trans}}} \\
    &= \sqrt{v^2_{\text{trans}}(\cot^2{i} + 1)} \\
    &= 4.74d\sqrt{\cot^2{i} + 1}\sqrt{\mu^2_{\delta} + \mu^2_{\alpha}\cos^2\delta}
\end{align}
where the declination $\delta$ is measured in radians. This was iterated 10000 times over the sampling parameters $d$ and $i$ and the result was converted to the Local Standard of Rest from SSB using the method given by \citet{McMillan2017}. We arrive at a value for the velocity of \psr{} of \vtot $\kms$, assuming YMW16. We have also performed this calculation for all DNSs with measured proper motions; these are reported in Table \ref{tab:PSRs}.
Although population sizes are relatively small, calculations suggest two different velocity environments with an overall upward trend with companion mass among DNSs. The two red points in Figure \ref{fig:velocity} represent PSRs B1534$+$12 and B1913$+$16. These pulsar binaries are thought to have formed in an asymmetric SN given their estimated kick velocities \citep{Tauris2017}. This hints at a distinct population divide to several other DNS systems such as J1756$-$2251 and J0737$-$3039 (both in blue). At its estimated median LSR velocity, \psr{} appears to be in the latter group, following the currently observed upward trend with respect to $m_{\text{c}}$. The uncertainties in its velocity are still somewhat too large to draw definitive conclusions.

We now calculate the kinematic contributions for $\dot{P}$. These are given by the second derivative of the line-of-sight distance from the pulsar to the Earth, or the first derivative of the Doppler factor. Using calculations in \citet{Stovall2019}, we obtain three main contributions: $8.28 \times 10^{-21}$ for the Shklovskii effect \citep{Shklovskii1970}, $-1.23 \times 10^{-21}$ for the difference in rotational accelerations between the Solar System and the pulsar, projected along the direction between the two, and $-1.09 \times 10^{-21}$ for the difference in vertical accelerations between the Solar System and the pulsar, projected along this same direction. The total correction to the spin period is then $5.97 \times 10^{-21}$.
Subtracting this from $\dot{P}$, we find an intrinsic $\dot{P}$ of $4.35 \pm 0.09 \times 10^{-20}$, and values for the pulsar characteristics as described in Table \ref{tab:1829_params}.

The equivalent kinematic contributions to $\dot{P}_{\text{b}}$ are $2.0 \times \, 10^{-14}$ (Shklovskii), $-3.0 \times 10^{-15}$ (rotation acceleration) and $-2.6 \times 10^{-15}$ (vertical acceleration difference). The total predicted kinematic contribution to $\dot{P}_{\text{b}}$ is then $+1.5\, \pm\, 0.2 \times 10^{-14}$. If we fit for $\dot{P}_{\text{b}}$ in DDH, we obtain  $-2.9\, \pm \,1.1 \times 10^{-14}$ ($1\,\sigma$). With the DDGR model, we can fit for a contribution to $\dot{P}_{\text{b}}$ that is in addition to the GR prediction for the masses of the system (this parameter is known as XPBDOT in \textsc{tempo2}). Fitting for this, we find the non-GR contribution of $\dot{P}^{\rm X}_{\text{b}}$ to be $-2.3\, \pm \, 1.1\times 10^{-14}$ ($1\,\sigma$). The difference between the prediction and observation is therefore $-3.8 \, \pm \, 1.1 \times \, 10^{-14}$, which is more than $3\,\sigma$ significant. As for the anomalous value of $\dot{x}$, this could be caused by systematics in our data, but if the effect is real, there may be some nearby mass accelerating the system. Continued timing will be necessary in order to verify this.

\subsection{Binary evolution} \label{sec:binaryevolution}

Under the assumption that the total space velocity for \psr{} is not entirely in the radial direction, so that our Monte Carlo approximation for $v_{\text{trans}}$ holds, the relatively low magnitude of the total proper motion ($\mu_{\text{tot}}$) when compared with other DNS systems supports a formation scenario for the companion NS that involves a low-kick supernova, which would be expected from a symmetric event \citep{Hills1983, Tauris2015}. This is further supported by its observed low eccentricity when compared with other Galactic DNSs \citep[see e.g.][]{Tauris2017}. The second SN could have been through either a rapid electron capture (EC) onto an O-Ne-Mg core or an ultra-stripped iron core-collapse (FeCC) scenario \citep{Tauris2013}. Most of these systems have companion masses that are not consistent with the narrow mass range predicted for formation in a ECSN. Since FeCCSNe have been shown to produce NS masses of $\sim 1.1 - 1.8\msun$ \citep{Tauris2015}, they are, according to current knowledge, the most likely formation channel for these low-kick systems. This evolutionary pathway is believed to be similar to the systems containing PSRs J0453$+$1559, J0737$-$3039, J1756$-$2251 and J1946$+$2052 based on these arguments.

The eccentricity of the orbit is inconsistent with the asymmetric SN pathway described in some models which relate eccentricity to the orbital separation for symmetric and asymmetric SN \citep[e.g.][]{Fryer1997, Willems2004, Tauris2017}.

\begin{table*}
	\centering
	\caption{Parameters for various DNS systems in which the pulsar is the recycled NS. This list does not include systems in globular clusters, which were likely formed via exchange encounters.\label{tab:PSRs}}
	\begin{tabular}{lccccccc}
		\hline
\multicolumn{1}{c}{PSR$^*$} & \makecell{$P$ \\ (ms)} & \makecell{$\pb$ \\ (days)} & $e$ & \makecell{Companion \\ mass ($\msun$)} & \makecell{$\mu_{\text{tot}}$ \\ (mas yr$^{-1}$)} & $d$ (kpc)$^\times$ & \makecell{ $v^{\text{LSR}}$\\ ($\kms$)$^\dagger$} \\
\hline
J0453$+$1559$^1$ & 45.8 & 4.072 & 0.113 & 1.174(4) & 7.997 & 0.52 & 29$^{+44}_{-19}$ \\ [4pt]
J0509$+$3801$^2$ & 76.5 & 0.380 & 0.586 & 1.46(8) & -- & $7.08$ & -- \\ [4pt]
J0737$-$3039A$^3$ & 22.7 & 0.102 & 0.088 & 1.2489(7) & 3.885 & 1.17 & 55$^{+86}_{-36}$ \\ [4pt]
J1411$+$2551$^4$ & 62.4 & 2.615 & 0.169 & $>0.92$ & $\sim 12$ & $1.13$ & 85$^{+120}_{-51}$ \\ [4pt]
J1518$+$4904$^5$ & 40.9 & 8.634 & 0.249 & 1.05$^{+1.21}_{-0.11}$ & 8.512 & 0.96 & 36$^{+55}_{-22}$ \\ [4pt]
B1534$+$12$^6$ & 37.9 & 0.421 & 0.274 & 1.3455(2) & 25.34 & 0.93 & 120$^{+184}_{-78}$ \\ [4pt]
J1753$-$2240$^7$ & 95.1 & 13.638 & 0.304 & -- & -- & 6.93 & -- \\ [4pt]
J1756$-$2251$^8$ & 28.5 & 0.320 & 0.181 & 1.230(7) & 5.928 & 0.95 & 42$^{+63}_{-25}$ \\ [4pt]
J1757$-$1854$^9$ & 21.5 & 0.183 & 0.606 & 1.3946(9) & -- & 19.6 & -- \\ [4pt]
J1811$-$1736$^{10}$ & 104.2 & 18.779 & 0.828 & $>0.93$ & -- & 10.16 & -- \\ [4pt]
\textbf{J1829$+$2456} & 41.1 & 1.176 & 0.139 & 1.306(7) & 9.560 & 0.91 & \vtot \\ [4pt]
J1913$+$1102$^{11}$ & 27.3 & 0.206 & 0.090 & 1.27(3) & 9.286 & 7.14 & 112$^{+175}_{-73}$ \\ [4pt]
B1913$+$16$^{12}$ & 59.0 & 0.323 & 0.617 & 1.389(1) & 1.404 & 5.25 & 157$^{+242}_{-100}$ \\ [4pt]
J1930$-$1852$^{13}$ & 185.5 & 45.060 & 0.399 & $>1.30$ & -- & 2.48 & -- \\ [4pt]
J1946$+$2052$^{14}$ & 16.9 & 0.078 & 0.064 & $>1.18$ & -- & 3.51 & -- \\ [4pt]
\hline
\end{tabular}
\flushleft
$^*$References: (1) \citet{Martinez2015}, (2) \citet{Lynch2018}, (3) \citet{Kramer2006}, (4) \citet{Martinez2017}, (5) \citet{Janssen2008}, (6) \citet{B1534}, (7) \citet{Keith2009}, (8) \citet{Ferdman2014}, (9) \citet{Cameron2018}, (10) \citet{Corongiu2007}, (11) \citet{Ferdman2020}, (12) \citet{Weisberg2005}, (13) \citet{Swiggum2015}, (14) \citet{Stovall2018}. \\
$^\times$Distances used were derived from the YMW16 Galactic free electron distribution model \citep{Yao2017} with DMs found using the ATNF Pulsar Catalogue \citep{Manchester2005} except in the case of PSR~J1756$-$2251, where the distance is given by \citet{Ferdman2014} \\
$^\dagger$Median $v^{\text{LSR}}$ and 2$\sigma$ (95\% confidence level) errors were calculated using the Monte-Carlo method described in section \ref{sec:velocities} and rounded to the nearest integer.
\end{table*}

\section{Conclusions}

We have presented an updated timing solution for the \psr{} DNS system. We have made precise measurements of both the pulsar and companion mass, finding them to be of similar mass, and have precisely determined the proper motion of the system. This implies a low transverse peculiar velocity. The eccentricity, velocity, and system masses, all found through timing, are similar to the evolutionary models of PSRs J0453+1559, J0737$-$3039, J1756$-$2251 and J1946$+$2052, in which the second-formed NS was very likely formed in a symmetric, low-kick supernova following a short-duration mass accretion process.

Given the small number of well measured NS masses and proper motions in DNS systems, the current measurements are important additions to our knowledge of the characteristics of the population. It is becoming increasingly apparent that a significant majority of the DNSs population in our Galaxy (6 out of 9 systems with $v^{\rm LSR}$ estimates; see Table~\ref{tab:PSRs}) has formed in similar small-kick events, with only two systems (PSRs~B1913$+$16, B1534$+$12 and J1518$+$4904) having strong associated kicks.

%This presents an explanation for the fact that the overwhelming majority of DNSs in our Galaxy are found at low Galactic heights; however, it conflicts with findings of locations of short gamma-ray bursts, which have a broad distribution of distances from the centres of their host galaxies, implicit of larger kick velocity scenarios. An analysis of the reasons for this discrepancy does not fit here, but it is an important avenue of research

\section*{Acknowledgements}

The authors wish to thank T.~Tauris for many helpful discussions on binary evolution.
RDF and KAB wish to acknowledge funding from the Royal Astronomical Society Undergraduate Summer Bursary scheme. DRL and MAM acknowledge support from the NSF awards AAG-1616042, OIA-1458952 and
PHY-1430284.

\section*{Data availability}

All data available from the author upon request.

\newpage

%%%%%%%%%%%%%%%%%%%%%%%%%%%%%%%%%%%%%%%%%%%%%%%%%%

%%%%%%%%%%%%%%%%%%%% REFERENCES %%%%%%%%%%%%%%%%%%

% The best way to enter references is to use BibTeX:

%\bibliographystyle{mnras}
%\bibliography{refs}

\begin{thebibliography}{}
\makeatletter
\relax
\def\mn@urlcharsother{\let\do\@makeother \do\$\do\&\do\#\do\^\do\_\do\%\do\~}
\def\mn@doi{\begingroup\mn@urlcharsother \@ifnextchar [ {\mn@doi@}
  {\mn@doi@[]}}
\def\mn@doi@[#1]#2{\def\@tempa{#1}\ifx\@tempa\@empty \href
  {http://dx.doi.org/#2} {doi:#2}\else \href {http://dx.doi.org/#2} {#1}\fi
  \endgroup}
\def\mn@eprint#1#2{\mn@eprint@#1:#2::\@nil}
\def\mn@eprint@arXiv#1{\href {http://arxiv.org/abs/#1} {{\tt arXiv:#1}}}
\def\mn@eprint@dblp#1{\href {http://dblp.uni-trier.de/rec/bibtex/#1.xml}
  {dblp:#1}}
\def\mn@eprint@#1:#2:#3:#4\@nil{\def\@tempa {#1}\def\@tempb {#2}\def\@tempc
  {#3}\ifx \@tempc \@empty \let \@tempc \@tempb \let \@tempb \@tempa \fi \ifx
  \@tempb \@empty \def\@tempb {arXiv}\fi \@ifundefined
  {mn@eprint@\@tempb}{\@tempb:\@tempc}{\expandafter \expandafter \csname
  mn@eprint@\@tempb\endcsname \expandafter{\@tempc}}}

\bibitem[\protect\citeauthoryear{{Antoniadis} et~al.,}{{Antoniadis}
  et~al.}{2013}]{Antoniadis2013}
{Antoniadis} J.,  et~al., 2013, \mn@doi [Science] {10.1126/science.1233232},
  340, 448

\bibitem[\protect\citeauthoryear{{Archibald} et~al.,}{{Archibald}
  et~al.}{2018}]{Archibald2018}
{Archibald} A.~M.,  et~al., 2018, \mn@doi [Nature] {10.1038/s41586-018-0265-1},
  559, 73

\bibitem[\protect\citeauthoryear{Arzoumanian et~al.,}{Arzoumanian
  et~al.}{2018}]{Arzoumanian2018}
Arzoumanian Z.,  et~al., 2018, \mn@doi [The Astrophysical Journal Supplement
  Series] {10.3847/1538-4365/aab5b0}, 235, 37

\bibitem[\protect\citeauthoryear{{Cameron} et~al.,}{{Cameron}
  et~al.}{2018}]{Cameron2018}
{Cameron} A.~D.,  et~al., 2018, \mn@doi [\mnras] {10.1093/mnrasl/sly003}, 475,
  L57

\bibitem[\protect\citeauthoryear{Champion, Lorimer, McLaughlin, Cordes,
  Arzoumanian, Weisberg  \& Taylor}{Champion et~al.}{2004}]{Champion2004}
Champion D.~J.,  Lorimer D.~R.,  McLaughlin M.~A.,  Cordes J.~M.,  Arzoumanian
  Z.,  Weisberg J.~M.,   Taylor J.~H.,  2004, \mn@doi [Monthly Notices of the
  Royal Astronomical Society]
  {http://dx.doi.org/10.1111/j.1365-2966.2004.07862.x}, 350, 61

\bibitem[\protect\citeauthoryear{Champion et al.}{2005}]{Champion2005} Champion D.~J., Lorimer D.~R., McLaughlin M.~A., Xilouris K.~M., Arzoumanian Z., Freire P.~C.~C., Lommen A.~N., et al., 2005, \mn@doi [Monthly Notices of the Royal Astronomical Society]{10.1111/j.1365-2966.2005.09499.x}, 363, 929


\bibitem[\protect\citeauthoryear{Cordes \& Lazio}{Cordes \&
  Lazio}{2003}]{Cordes2003}
Cordes J.~M.,  Lazio T. J.~W.,  2003 (\mn@eprint {arXiv} {astro-ph/0301598})

\bibitem[\protect\citeauthoryear{{Corongiu}, {Kramer}, {Stappers}, {Lyne},
  {Jessner}, {Possenti}, {D'Amico}  \& {L{\"o}hmer}}{{Corongiu}
  et~al.}{2007}]{Corongiu2007}
{Corongiu} A.,  {Kramer} M.,  {Stappers} B.~W.,  {Lyne} A.~G.,  {Jessner} A.,
  {Possenti} A.,  {D'Amico} N.,   {L{\"o}hmer} O.,  2007, \mn@doi [Astronomy
  and Astrophysics] {10.1051/0004-6361:20054385}, 462, 703

\bibitem[\protect\citeauthoryear{{Damour} \& {Deruelle}}{{Damour} \&
  {Deruelle}}{1985}]{DD1985}
{Damour} T.,  {Deruelle} N.,  1985, Ann. Inst. Henri Poincar{\'e} Phys.
  Th{\'e}or, 43, 107

\bibitem[\protect\citeauthoryear{{Damour} \& {Deruelle}}{{Damour} \&
  {Deruelle}}{1986}]{DD1986}
{Damour} T.,  {Deruelle} N.,  1986, Ann. Inst. Henri Poincar{\'e} Phys.
  Th{\'e}or, 44, 263

\bibitem[\protect\citeauthoryear{Damour \& Taylor}{1992}]{DT92} Damour T., Taylor J.~H., 1992, PhRvD, 45, 1840

\bibitem[\protect\citeauthoryear{{Damour} \& {Esposito-Far\`ese}}{{Damour} \&
  {Esposito-Far\`ese}}{1992}]{Damour1992}
{Damour} T.,  {Esposito-Far\`ese} G.,  1992, \mn@doi [Classical and Quantum
  Gravity] {10.1088/0264-9381/9/9/015}, 9, 2093

\bibitem[\protect\citeauthoryear{Damour \& Esposito-Far\`ese}{Damour \&
  Esposito-Far\`ese}{1993}]{Damour1993}
Damour T.,  Esposito-Far\`ese G.,  1993, Phys. Rev. Lett., 70, 2220

\bibitem[\protect\citeauthoryear{Damour \& Esposito-Far\`ese}{Damour \&
  Esposito-Far\`ese}{1996}]{Damour1996}
Damour T.,  Esposito-Far\`ese G.,  1996, \mn@doi [Physical Review D]
  {10.1103/physrevd.54.1474}, 54, 1474–1491

\bibitem[\protect\citeauthoryear{Damour \& Taylor}{Damour \&
  Taylor}{1992}]{dt92}
Damour T.,  Taylor J.~H.,  1992, \mn@doi [Phys. Rev. D]
  {10.1103/PhysRevD.45.1840}, 45, 1840

\bibitem[\protect\citeauthoryear{Edwards, Hobbs  \& Manchester}{Edwards
  et~al.}{2006}]{Edwards2006}
Edwards R.~T.,  Hobbs G.~B.,   Manchester R.~N.,  2006, \mn@doi [Monthly
  Notices of the Royal Astronomical Society]
  {10.1111/j.1365-2966.2006.10870.x}, 372, 1549–1574

\bibitem[\protect\citeauthoryear{Ferdman et~al.,}{Ferdman
  et~al.}{2013}]{Ferdman2013}
Ferdman R.~D.,  et~al., 2013, The Astrophysical Journal, 767, 85

\bibitem[\protect\citeauthoryear{{Ferdman} et~al.,}{{Ferdman}
  et~al.}{2014}]{Ferdman2014}
{Ferdman} R.~D.,  et~al., 2014, \mn@doi [Monthly Notices of the Royal
  Astronomical Society] {10.1093/mnras/stu1223}, 443, 2183

\bibitem[\protect\citeauthoryear{Ferdman et~al.,}{Ferdman
  et~al.}{2020}]{Ferdman2020}
Ferdman R.~D.,  et~al., 2020, \mn@doi [Nature] {10.1038/s41586-020-2439-x},
  583, 211

\bibitem[\protect\citeauthoryear{Folkner, Park  \& Jacobson}{Folkner
  et~al.}{2016}]{DE435}
Folkner W.~M.,  Park R.~S.,   Jacobson R.~A.,  2016, JPL IOM 392R-16-003

\bibitem[\protect\citeauthoryear{{Fonseca}, {Stairs}  \& {Thorsett}}{{Fonseca}
  et~al.}{2014}]{B1534}
{Fonseca} E.,  {Stairs} I.~H.,   {Thorsett} S.~E.,  2014, \mn@doi [The
  Astrophysical Journal] {10.1088/0004-637X/787/1/82}, 787, 82

\bibitem[\protect\citeauthoryear{{Freire} \& {Wex}}{{Freire} \&
  {Wex}}{2010}]{DDH}
{Freire} P.~C.~C., {Wex} N.,  2010, \mn@doi [Monthly Notices of the Royal Astronomical Society]
  {10.1111/j.1365-2966.2010.17319.x}, 409, 199

\bibitem[\protect\citeauthoryear{{Freire} et~al.}{{Freire}
  et~al.}{2012}]{Freire2012}
Freire P.~C.~C., Wex N., Esposito-Far{\`e}se G., Verbiest J.~P.~W., Bailes M., Jacoby B.~A., Kramer M., et~al.,  2012, \mn@doi [Monthly Notices of the Royal Astronomical Society]
  {10.1111/j.1365-2966.2012.21253.x}, 423, 3328

\bibitem[\protect\citeauthoryear{{Fryer} \& {Kalogera}}{{Fryer} \&
  {Kalogera}}{1997}]{Fryer1997}
{Fryer} C.,  {Kalogera} V.,  1997, \mn@doi [The Astrophysical Journal]
  {10.1086/304772}, 489, 244

\bibitem[\protect\citeauthoryear{{Gehrels}, {Macomb}, {Bertsch}, {Thompson}  \&
  {Hartman}}{{Gehrels} et~al.}{2000}]{Gehrels2000}
{Gehrels} N.,  {Macomb} D.~J.,  {Bertsch} D.~L.,  {Thompson} D.~J.,   {Hartman}
  R.~C.,  2000, \mn@doi [Nature] {10.1038/35006001}, 404, 363

\bibitem[\protect\citeauthoryear{{Grenier}}{{Grenier}}{2000}]{Grenier2000}
{Grenier} I.~A.,  2000, Astronomy and Astrophysics, 364, L93

\bibitem[\protect\citeauthoryear{{Guinot}}{{Guinot}}{1988}]{Guinot1988}
{Guinot} B.,  1988, Astronomy and Astrophysics, 192, 370

\bibitem[\protect\citeauthoryear{{Hills}}{{Hills}}{1983}]{Hills1983}
{Hills} J.~G.,  1983, \mn@doi [The Astrophysical Journal] {10.1086/160871},
  \href {https://ui.adsabs.harvard.edu/abs/1983ApJ...267..322H} {267, 322}

\bibitem[\protect\citeauthoryear{Hobbs, Edwards  \& Manchester}{Hobbs
  et~al.}{2006}]{Hobbs2006}
Hobbs G.~B.,  Edwards R.~T.,   Manchester R.~N.,  2006, \mn@doi [Monthly
  Notices of the Royal Astronomical Society]
  {10.1111/j.1365-2966.2006.10302.x}, 369, 655–672

\bibitem[\protect\citeauthoryear{{Hulse} \& {Taylor}}{{Hulse} \&
  {Taylor}}{1975}]{Hulse1975}
{Hulse} R.~A.,  {Taylor} J.~H.,  1975, The Astrophysical Journal, 456, L51

\bibitem[\protect\citeauthoryear{{Janssen}, {Stappers}, {Kramer}, {Nice},
  {Jessner}, {Cognard}  \& {Purver}}{{Janssen} et~al.}{2008}]{Janssen2008}
{Janssen} G.~H.,  {Stappers} B.~W.,  {Kramer} M.,  {Nice} D.~J.,  {Jessner} A.,
   {Cognard} I.,   {Purver} M.~B.,  2008, \mn@doi [Astronomy and Astrophysics]
  {10.1051/0004-6361:200810076}, 490, 753

\bibitem[\protect\citeauthoryear{Keith, Kramer, Lyne, Eatough, Stairs,
  Possenti, Camilo  \& Manchester}{Keith et~al.}{2009}]{Keith2009}
Keith M.~J.,  Kramer M.,  Lyne A.~G.,  Eatough R.~P.,  Stairs I.~H.,  Possenti
  A.,  Camilo F.,   Manchester R.~N.,  2009, \mn@doi [Monthly Notices of the
  Royal Astronomical Society] {10.1111/j.1365-2966.2008.14234.x}, 393, 623

\bibitem[\protect\citeauthoryear{{Kramer} et~al.,}{{Kramer}
  et~al.}{2006}]{Kramer2006}
{Kramer} M.,  et~al., 2006, \mn@doi [Science] {10.1126/science.1132305}, 314,
  97

\bibitem[\protect\citeauthoryear{Krishnan et~al.,}{Krishnan
  et~al.}{2020}]{Krishnan2020}
Krishnan V.~V.,  et~al., 2020, \mn@doi [Science] {10.1126/science.aax7007},
  367, 577

\bibitem[\protect\citeauthoryear{{Lorimer, D. R. and Kramer, M.}}{{Lorimer, D.
  R. and Kramer, M.}}{2005}]{Lorimer2004}
{Lorimer, D. R. and Kramer, M.} 2005, {Handbook of Pulsar Astronomy}.
Cambridge University Press

\bibitem[\protect\citeauthoryear{Lynch et~al.,}{Lynch et~al.}{2018}]{Lynch2018}
Lynch R.~S.,  et~al., 2018, \mn@doi [The Astrophysical Journal]
  {10.3847/1538-4357/aabf8a}, 859, 93

\bibitem[\protect\citeauthoryear{{Manchester}, {Hobbs}, {Teoh}  \&
  {Hobbs}}{{Manchester} et~al.}{2005}]{Manchester2005}
{Manchester} R.~N.,  {Hobbs} G.~B.,  {Teoh} A.,   {Hobbs} M.,  2005, \mn@doi
  [The Astronomical Journal] {10.1086/428488}, 129, 1993

\bibitem[\protect\citeauthoryear{Martinez et~al.,}{Martinez
  et~al.}{2015}]{Martinez2015}
Martinez J.~G.,  et~al., 2015, \mn@doi [The Astrophysical Journal]
  {https://doi.org/10.1088/0004-637X/812/2/143}, 812

\bibitem[\protect\citeauthoryear{Martinez et~al.,}{Martinez
  et~al.}{2017}]{Martinez2017}
Martinez J.~G.,  et~al., 2017, \mn@doi [The Astrophysical Journal]
  {10.3847/2041-8213/aa9d87}, 851, L29

\bibitem[\protect\citeauthoryear{McMillan}{McMillan}{2017}]{McMillan2017}
McMillan P.~J.,  2017, \mn@doi [Monthly Notices of the Royal Astronomical
  Society] {10.1093/mnras/stw2759}, 465, 76

\bibitem[\protect\citeauthoryear{{Nomoto}}{{Nomoto}}{1984}]{Nomoto1984}
{Nomoto} K.,  1984, \mn@doi [The Astrophysical Journal] {10.1086/161749}, 277,
  791

\bibitem[\protect\citeauthoryear{Ridolfi, Freire, Gupta  \& Ransom}{Ridolfi
  et~al.}{2019}]{Ridolfi2019}
Ridolfi A.,  Freire P. C.~C.,  Gupta Y.,   Ransom S.~M.,  2019, \mn@doi
  [Monthly Notices of the Royal Astronomical Society] {10.1093/mnras/stz2645},
  490, 3860

\bibitem[\protect\citeauthoryear{{Shklovskii}}{{Shklovskii}}{1970}]{Shklovskii1970}
{Shklovskii} I.~S.,  1970, Soviet Astronomy, 13, 562

\bibitem[\protect\citeauthoryear{Stairs, Thorsett, Taylor  \& Wolszczan}{Stairs
  et~al.}{2002}]{Stairs2002}
Stairs I.~H.,  Thorsett S.~E.,  Taylor J.~H.,   Wolszczan A.,  2002, The
  Astrophysical Journal, 581, 501

\bibitem[\protect\citeauthoryear{{Stovall} et~al.,}{{Stovall}
  et~al.}{2018}]{Stovall2018}
{Stovall} K.,  et~al., 2018, \mn@doi [The Astrophysical Journal Letters]
  {10.3847/2041-8213/aaad06}, 854, L22

\bibitem[\protect\citeauthoryear{Stovall et~al.,}{Stovall
  et~al.}{2019}]{Stovall2019}
Stovall K.,  et~al., 2019, \mn@doi [The Astrophysical Journal]
  {10.3847/1538-4357/aaf37d}, 870, 74

\bibitem[\protect\citeauthoryear{{Swiggum} et~al.,}{{Swiggum}
  et~al.}{2015}]{Swiggum2015}
{Swiggum} J.~K.,  et~al., 2015, \mn@doi [The Astrophysical Journal]
  {10.1088/0004-637X/805/2/156}, 805, 156

\bibitem[\protect\citeauthoryear{{Tauris}
   et~al.}{2013}]{Tauris2013} Tauris T.~M., Langer N., Moriya T.~J., Podsiadlowski P., Yoon S.-C., Blinnikov S.~I., 2013, \mn@doi [ Astrophysical Journal Letters ]{10.1088/2041-8205/778/2/L23}, 778, L23

\bibitem[\protect\citeauthoryear{{Tauris}, {Langer}  \&
  {Podsiadlowski}}{{Tauris} et~al.}{2015}]{Tauris2015}
{Tauris} T.~M.,  {Langer} N.,   {Podsiadlowski} P.,  2015, \mn@doi [Monthly
  Notices of the Royal Astronomical Society] {10.1093/mnras/stv990}, 451, 2123

\bibitem[\protect\citeauthoryear{Tauris et~al.,}{Tauris
  et~al.}{2017}]{Tauris2017}
Tauris T.~M.,  et~al., 2017, \mn@doi [The Astrophysical Journal]
  {https://doi.org/10.3847/1538-4357/aa7e89}, 846

\bibitem[\protect\citeauthoryear{{Tauris} \& {Janka}}{{Tauris} \&
  {Janka}}{2019}]{Tauris2019}
{Tauris} T.~M.,  {Janka} H.-T.,  2019, \mn@doi [The Astrophysical Journal
  Letters] {10.3847/2041-8213/ab5642}, 886, L20

\bibitem[\protect\citeauthoryear{{Taylor}}{{Taylor}}{1992}]{Taylor1992}
{Taylor} J.~H.,  1992, \mn@doi [Philosophical Transactions of the Royal Society
  of London Series A] {10.1098/rsta.1992.0088}, \href
  {https://ui.adsabs.harvard.edu/abs/1992RSPTA.341..117T} {341, 117}

\bibitem[\protect\citeauthoryear{{Taylor} \& {Weisberg}}{{Taylor} \&
  {Weisberg}}{1989}]{DDGR1989}
{Taylor} J.~H.,  {Weisberg} J.~M.,  1989, \mn@doi [The Astrophysical Journal]
  {10.1086/167917}, 345, 434

\bibitem[\protect\citeauthoryear{Voisin, Cognard, Freire, Wex, Guillemot,
  Desvignes, Kramer  \& Theureau}{Voisin et~al.}{2020}]{Voisin2020}
Voisin G.,  Cognard I.,  Freire P. C.~C.,  Wex N.,  Guillemot L.,  Desvignes
  G.,  Kramer M.,   Theureau G.,  2020, \mn@doi [A\&A]
  {10.1051/0004-6361/202038104}, 638, A24

\bibitem[\protect\citeauthoryear{Wang, Podsiadlowski  \& Han}{Wang
  et~al.}{2017}]{Wang2017}
Wang B.,  Podsiadlowski P.,   Han Z.,  2017, Monthly Notices of the Royal
  Astronomical Society, 472, 1593

\bibitem[\protect\citeauthoryear{Weisberg \& Huang}{Weisberg \&
  Huang}{2016}]{Weisberg2016}
Weisberg J.~M.,  Huang Y.,  2016, \mn@doi [The Astrophysical Journal]
  {10.3847/0004-637x/829/1/55}, 829, 55

\bibitem[\protect\citeauthoryear{{Weisberg} \& {Taylor}}{{Weisberg} \&
  {Taylor}}{2005}]{Weisberg2005}
{Weisberg} J.~M.,  {Taylor} J.~H.,  2005, in {Rasio} F.~A.,  {Stairs} I.~H.,
  eds,  Astronomical Society of the Pacific Conference Series Vol. 328, Binary
  Radio Pulsars. p.~25

\bibitem[\protect\citeauthoryear{Wex, Kalogera  \& Kramer}{Wex
  et~al.}{2000}]{Wex2000}
Wex N.,  Kalogera V.,   Kramer M.,  2000, \mn@doi [The Astrophysical Journal]
  {10.1086/308148}, 528, 401

\bibitem[\protect\citeauthoryear{{Willems} \& {Kalogera}}{{Willems} \&
  {Kalogera}}{2004}]{Willems2004}
{Willems} B.,  {Kalogera} V.,  2004, \mn@doi [The Astrophysical Journal
  Letters] {10.1086/383200}, 603, L101

\bibitem[\protect\citeauthoryear{Yao, Manchester  \& Wang}{Yao
  et~al.}{2017}]{Yao2017}
Yao J.~M.,  Manchester R.~N.,   Wang N.,  2017, \mn@doi [The Astrophysical
  Journal] {10.3847/1538-4357/835/1/29}, 835, 29

\bibitem[\protect\citeauthoryear{{Zha}, {Leung}, {Suzuki}  \& {Nomoto}}{{Zha}
  et~al.}{2019}]{Zha2019}
{Zha} S.,  {Leung} S.-C.,  {Suzuki} T.,   {Nomoto} K.,  2019, \mn@doi [The
  Astrophysical Journal] {10.3847/1538-4357/ab4b4b}, 886, 22

\makeatother
\end{thebibliography}
% if your bibtex file is called example.bib

% Alternatively you could enter them by hand, like this:
% This method is tedious and prone to error if you have lots of references

%%%%%%%%%%%%%%%%%%%%%%%%%%%%%%%%%%%%%%%%%%%%%%%%%%

%%%%%%%%%%%%%%%%% APPENDICES %%%%%%%%%%%%%%%%%%%%%

%%%%%%%%%%%%%%%%%%%%%%%%%%%%%%%%%%%%%%%%%%%%%%%%%%

% Don't change these lines
\bsp    % typesetting comment
\label{lastpage}
\end{document}